\documentclass[a4paper,11pt]{article}

\usepackage{timecenter}

\usepackage{amsmath,amsfonts,amsthm,amstext,amssymb}
\usepackage{graphicx}
\usepackage[below]{placeins}
\usepackage{subfigure}
\usepackage{wrapfig}
\usepackage{times}

\makeatletter
\newcommand\figcaption{\def\@captype{figure}\caption}
\newcommand\tabcaption{\def\@captype{table}\caption}
\makeatother

\newtheorem{alg}{Algorithm}[section]

\newtheorem{Rem}{Remark}[section]

\DeclareGraphicsExtensions{.pdf,.pstex,.eps}

\begin{document}

\title{The INFATI Data}

\author{C.~S.~Jensen, H.~Lahrmann, S.~Pakalnis, and J.~Runge} 

\publication_history{July 2004, a \textsc{TimeCenter} Technical Report}

\trnumber{79}

\maketitle

\begin{abstract}
  
  The ability to perform meaningful empirical studies is of essence in
  research in spatio-temporal query processing. Such studies are often
  necessary to gain detailed insight into the functional and
  performance characteristics of proposals for new query processing
  techniques.
  We present a collection of spatio-temporal data, collected during an
  intelligent speed adaptation project, termed INFATI, in which some
  two dozen cars equipped with GPS receivers and logging equipment
  took part. We describe how the data was collected and how it was
  ``modified'' to afford the drivers some degree of anonymity.
  We also present the road network in which the cars were moving
  during data collection.
  The GPS data is publicly available for non-commercial purposes.  It
  is our hope that this resource will help the spatio-temporal
  research community in its efforts to develop new and better query
  processing techniques.

\end{abstract}

\section{Introduction}
\label{sec:intro}

Aspects of key computing and communication hardware technologies
continue to improve rapidly, some at sustained exponential rates.  The
advances in computing and communication combine with advances in
geo-positioning to enable a range of new, location-enabled, mobile
services. This entire development contributes to making research in
spatio-temporal data management more relevant than ever.

When developing new query processing techniques, prototype
implementation and subsequent rigorous empirical studies of central
functional and performance characteristics of the techniques are often
essential. Such studies may be the only or best means of gaining the
detailed insight necessary to guide the design process, and they may
be the only or best means of understanding the characteristics of the
final designs.

When subjecting query processing techniques to empirical study,
synthetic as well as real data play important roles. These kinds of
data have complimentary strengths and weaknesses.
Synthetic data are important for several reasons. First, a single real
data set is likely to capture only a specific type of use of the
technique under study. In order to test the technique under varying
types of conditions, synthetic data is useful. Second, synthetic data
generators offer controls that enable the generation of data sets with
specific properties, e.g., data sets with certain sizes and that
possess certain statistical properties. Synthetic data sets thus make
it possible to subject a techniques to a wide variety of conditions.
In contrast, real data are essential in guaranteeing that the
techniques under study are subjected to realistic conditions. With
synthetic data, there is generally no guarantee that the data
corresponds to any real-world application.

The literature offers descriptions of several synthetic-data
generators. In particular, a recent special issue of the IEEE Data
Engineering Bulletin contains papers that offer overviews of available
data generators and real data sets~\cite{jen03}.

In this special issue, Brinkhoff~\cite{bri03} surveys the generation
of data sets intended expressly for the testing of query processing
techniques underlying location-based services. Specifically, he covers
his own Network-based Generator~\cite{brinkhoff} and Kaufman et al.'s
City Simulator~\cite{kmj01}, both of which assume that the object
movement, from which the generated data result, is constrained to a
transportation network.

Also in this issue, Nascimento et al.~\cite{nas03} and Manolopoulos et
al.~\cite{man03} cover three data generators for moving objects that
differ from those covered by Brinkhoff in that they do not constrain
movement to a network. Stated briefly, GSTD~\cite{ascona00} generates
moving-point and moving-rectangle data. G-TERD~\cite{gterd} produces
sequences of raster images. Oporto~\cite{oporto} generates data
corresponding to fishing-at-sea scenarios.

Finally, Nascimento et al.\ cover several real data sets. Two data
sets contain animal-tracking data.  Another data set contains
hurricane tracking data. With less than a thousand data entries each,
these data sets are relatively small.  The data set most closely
related to the INFATI data contains data obtained from GPS receivers
attached to thirteen buses. Positions were sampled every 30 seconds
within a 24-hour interval, and the total number of entries is 28.617.
The sampling frequency in the INFATI data is much higher, and the
number of data entries is ca.~1.9 million.
Pfoser maintains a web page with pointers to real spatio-temporal
data~\cite{pfoser_data}.

The next section describes the general setting in which the INFATI GPS
log data were collected. Section~\ref{sec:gps} describes the GPS data,
including how some degree of driver-anonymity was ensured. Next,
Section~\ref{sec:roadnet} describes the road network in which the cars
were traveling when the GPS data was collected.
Section~\ref{sec:download} details how to download the data and
documentation. A final section offers acknowledgments.

\section{Background Information}

\begin{wrapfigure}{l}[0pt]{325pt}
  \vspace*{-24pt}
  \begin{center}
    \subfigure[Within the Speed Limit]{
    \epsfxsize=150pt
    \epsfbox{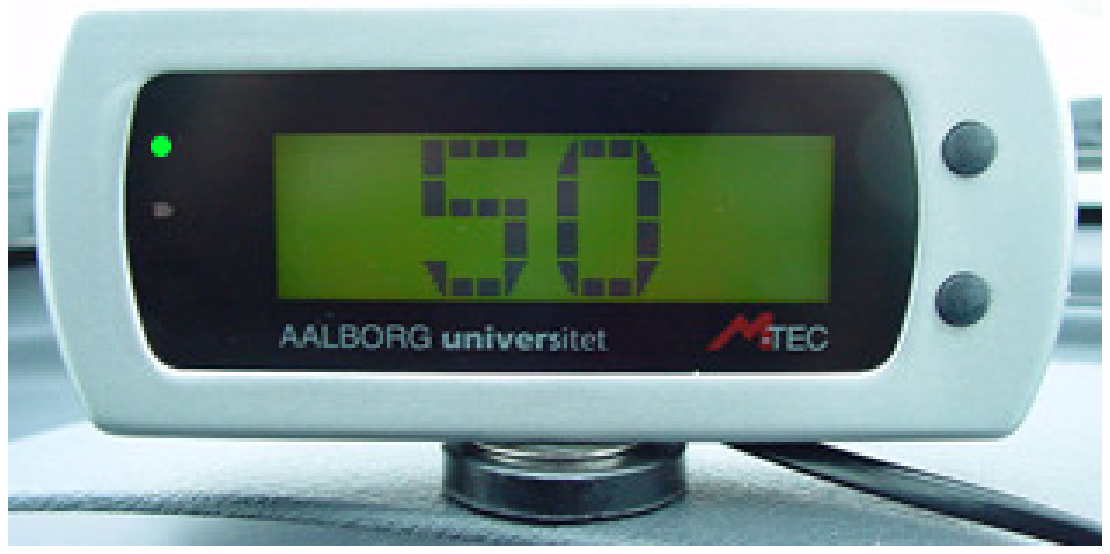}
    }
    \hspace{5pt}
    \subfigure[Exceeding the Speed Limit]{
    \epsfxsize=150pt
    \epsfbox{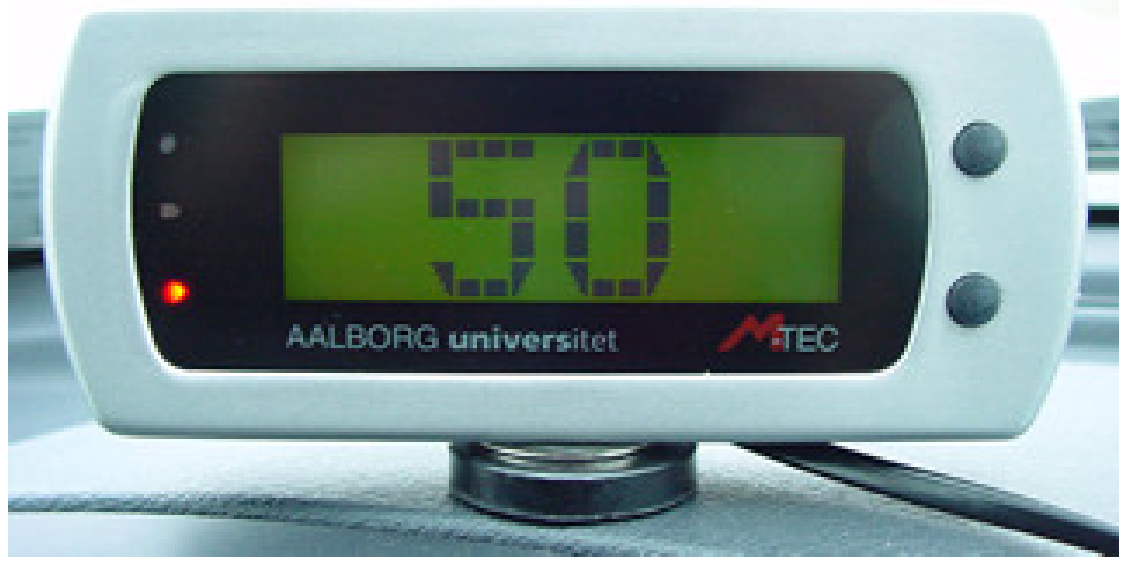}
    }
  \end{center}
  \vspace*{-12pt}
    \caption{The INFATI In-Vehicle Display}
    \label{fig:display}
\end{wrapfigure}

The INFATI data derive from the INFATI Project~\cite{www:infati}, an
intelligent speed adaptation project carried out by a team of
researchers at department of Development and Planning, Aalborg
University, that also included participants from the companies Sven
Allan Jensen and M-Tec.
The main purpose of the project was to investigate driver response to
alerts issued by a device installed in the car. This device
continuously displays the current speed. When the speed is below the
speed limit, the screen features a green light (see
Figure~\ref{fig:display}(a)). When the speed exceeds the limit, the
green light is replaced by a flashing red light (see
Figure~\ref{fig:display}(b)) and the display of the current speed also
flashes. In addition, a female voice announces the speed limit, adding
``you are driving too fast'' (in Danish).

A total of 24 distinct test cars and families participated in the
INFATI project's intelligent speed adaptation experiment. The cars
were divided into two teams, \emph{Team-1} and \emph{Team-2}. The
INFATI data contains GPS log-data from 11 cars from \emph{Team-1}.
This data was collected during December 2000 and January 2001. The
INFATI data contains data from 9 cars in \emph{Team-2}. This data was
collected during February and March 2001.

The remaining
4 cars were excluded for varying reasons. All cars were driving in the
municipality of Aalborg, which includes the city of Aalborg, suburbs,
and some neighboring towns. Section~\ref{sec:roadnet} describes the
road network of this area in some detail.
 
In addition to the display, each car was equipped with a Global
Positioning System (GPS)~\cite{www:gps} receiver and a small
custom-built computer. For more than a month, the movement of each car
was registered in the car's database. When a car was moving, its GPS
position was sampled every second. The GPS positions were stored in
the Universal Transverse Mercator (UTM 32) format. No sampling was
performed when a car was parked. Additional information about the
experiment can be found on the INFATI web site~\cite{www:infati}.

\section{GPS Positions of Cars}
\label{sec:gps}

For each car that delivered data, the INFATI data contains one file
with GPS log data. This section first describes the contents of the 20
resulting files, then describes the data-removal procedure that was
applied in order to introduce some degree of driver privacy
protection.

\subsection{GPS Log Files}

The GPS log data files are named as follows:
\emph{team$\mathrm{T}$\_car$\mathrm{C}$\_no\_home.txt}, where
$\mathrm{T}$ represents the number of the team and $\mathrm{C}$
represents the unique car identifier.  For example,
\emph{team$1$\_car$3$\_no\_home.txt} is the file for car number 3 in
\emph{Team-1}. The two teams were active in non-overlapping time
periods.  Statistics about the cars are provided in
Tables~\ref{fig:teams}(a)~and~(b).

\begin{figure}[!htb]
  \begin{center}
    \subfigure[\emph{Team-1} Statistics]{
\begin{tabular}{l|rlll}
 Car id          &     Records &                Earliest date &         Latest date     \\
\hline
        1   &     47055 & 22-Dec-00 & 22-Jan-01 \\
        2     &   79607 & 06-Dec-00 & 29-Jan-01 \\
          3    &    73189       & 07-Dec-00  & 25-Jan-01 \\
      4    &    14291   & 08-Dec-00  & 31-Dec-00 \\
        6       & 30361        & 21-Dec-00  & 30-Jan-01 \\
        7      &  37438 & 22-Dec-00  & 23-Jan-01 \\
         8      &  46290         & 22-Dec-00  & 22-Jan-01 \\
         9      &  87785        & 02-Jan-01  & 30-Jan-01 \\
        10     &   63536        & 02-Jan-01  & 30-Jan-01 \\
       11      &  86699 & 25-Dec-00  & 10-Jan-01 \\
          12      &  117873  &08-Dec-00  & 29-Jan-01 \\
         \multicolumn{ 2 }{  r }{Total:     684124} &    & \\
\end{tabular}
    }
    \hspace{5pt}
    \subfigure[\emph{Team-2} Statistics]{
\begin{tabular}{l|rlll}
 Car id          &     Records &                Earliest date &         Latest date     \\
\hline
          1 &      264721& 11-Feb-01 &26-Mar-01\\
         2   &     85549& 05-Feb-01 &26-Mar-01\\
         4   &    125476 &05-Feb-01 &26-Mar-01\\
         5     &  176477&03-Feb-01 &26-Mar-01\\
         6    &   113912 &14-Feb-01 &26-Mar-01\\
         8   &    163119 &05-Feb-01 &26-Mar-01\\
        10    &   100296 &07-Feb-01& 26-Mar-01\\
        11   &     63664& 06-Feb-01 &26-Mar-01\\
        12    &   117747& 07-Feb-01& 27-Mar-01\\
      \multicolumn{ 2 }{  r }{Total: 1210961} &    & \\
      \multicolumn{ 1 }{  r }{ }  &   &    & \\
      \multicolumn{ 1 }{  r }{ }  &   &    & \\
\end{tabular}
}
\end{center}
  \vspace*{-18pt}
\tabcaption{Statistics on GPS Logs}
\label{fig:teams}
\end{figure}
The tables list the counts of GPS coordinates for a particular
\emph{Car id} and also give the time intervals, ranging from
\emph{Earliest date} to \emph{Latest date}, covered by the individual
cars.  Notice that car identifiers are unique only within teams, not
globally.

Next, Table~\ref{d:gps} describes the format of a GPS log data entry.
\begin{table}[!ht]
\begin{tabular}{p{1.4cm}|p{1cm}|p{12cm}}
Attribute &   Length        & Description \\
\hline
\emph{id}         & 12   & Entry identifier, unique within a team.\\
\emph{entryid}    & 14   & Identifier composed by the attributes: carid, rdate, and rtime.\\
\emph{carid}      &  2   & Car identifier, unique within a team. \\
\emph{driverid}   &  2   & Car driver identifier.\\
\emph{rdate}      &  6   & Date in the format \emph{DDMMYY} (where
\emph{D} denotes day, \emph{M} denotes month, and \emph{Y} denotes year).\\
\emph{rtime}      &  6   & Time in the format \emph{hhmmss} (where
\emph{h} denotes hours, \emph{m} denotes minutes, and \emph{s} denotes
seconds).  \\  
\emph{xcoord}     &  6   & X coordinate received from GPS receiver.  \\
\emph{ycoord}     &  7   & Y coordinate received from GPS receiver.  \\
\emph{mpx}        &  6   & Map-matched X coordinate.  \\
\emph{mpy}        &  7   & Map-matched Y coordinate.  \\
\emph{sat}        &  2   & The number of satellites used for
determining the current position of the car.\\
\emph{hdop}       &  2   & Horizontal dilution of precision.\\
\emph{maxspd}     &  3   & Speed limit on the road to which the car's
position is map-matched.\\
\emph{spd}        &  3   & Actual speed of the car.\\
\emph{strtcod}    &  4   & Street code of the street to which the
car's position is map-matched. \\
\end{tabular}
\caption{GPS Log Data Entry Format}
\label{d:gps}
\end{table}
A few comments are in order. Attribute \emph{carid} is unique only
within a team (recall Tables~\ref{fig:teams}(a)~and~(b)).
However, the two teams were composed of different cars, meaning that
no single car participated in both teams. A car has one or more
drivers. Attributes \emph{rdate} and \emph{rtime} record the date and
time when an entry was measured---in standard temporal database terms,
they denote valid time.
Attribute \emph{entryid} is a concatenation of the \emph{carid},
\emph{rdate}, and \emph{rtime} of an entry. As the granularity of the
\emph{rtime} attribute is second, and as we sample with the frequency
of one second, one may expect \emph{entryid} to be unique within a
team and a file. However, it turns out that there does exist entries
for the same car, date, and second.
For attribute \emph{strtcod}, the exceptional value ``$-9$'' indicates
that the GPS position in an entry could not be mapped to any street.

Table~\ref{sample:gps} contains a few GPS log data entries.
\begin{table}[!ht] \footnotesize
\begin{tabular}{@{\hspace{0.2cm}} l @{\hspace{0.2cm}} l@{\hspace{0.2cm}}l@{\hspace{0.2cm}}l@{\hspace{0.2cm}}l@{\hspace{0.2cm}}l@{\hspace{0.2cm}}l@{\hspace{0.2cm}}l@{\hspace{0.2cm}}l@{\hspace{0.2cm}}l@{\hspace{0.2cm}}l@{\hspace{0.2cm}}l@{\hspace{0.2cm}}l@{\hspace{0.2cm}}l@{\hspace{0.2cm}}l@{\hspace{0.2cm}}}

\emph{id} & \emph{entryid} & \emph{carid} & \emph{driverid} & \emph{rdate} & \emph{rtime} & \emph{xcoord} & \emph{ycoord} &
\emph{mpx} & \emph{mpy} & \emph{sat} & \emph{hdop} & \emph{maxspd} & \emph{spd} & \emph{strtcod} \\
\hline
991  &12091200130310  &12    &0   &91200  &130310  &553570  &6315889  &553581  &6315886   &6   &1  &110  &101  &5490 \\ 
992  &12091200130311  &12    &0   &91200  &130311  &553562  &6315863  &553572  &6315859   &7   &1  &110  &101  &5490 \\
993  &12091200130312  &12    &0   &91200  &130312  &553554  &6315836  &553563  &6315833   &7   &1  &110  &101  &5490 \\
994  &12091200130313  &12    &0   &91200  &130313  &553547  &6315808  &553556  &6315806   &7   &1  &110  &100  &5490 \\
995  &12091200130314  &12    &0   &91200  &130314  &553541  &6315781  &553548  &6315779   &7   &1  &110  &100  &5490 \\
996  &12091200130315  &12    &0   &91200  &130315  &553534  &6315754  &553541  &6315752   &7   &1  &110  &101  &5490 \\
997  &12091200130316  &12    &0   &91200  &130316  &553528  &6315726  &553535  &6315725   &7   &1  &110  &101  &5490 \\
998  &12091200130317  &12    &0   &91200  &130317  &553523  &6315699  &553530  &6315697   &7   &1  &110  &101  &5490 \\
999  &12091200130318  &12    &0   &91200  &130318  &553518  &6315671  &553525  &6315670   &7   &1  &110  &101  &5490 \\
\end{tabular}
\caption{GPS Log Data Entries}
\label{sample:gps}
\end{table}
Observe that leading zeros are stripped from \emph{carid},
\emph{rdate}, and \emph{rtime}. However, values of the \emph{entryid}
attribute preserve leading zeros for each composing attribute, expect
\emph{carid}.

\begin{figure}[!ht]
\begin{center}
\includegraphics[angle=0,width=0.9\textwidth]{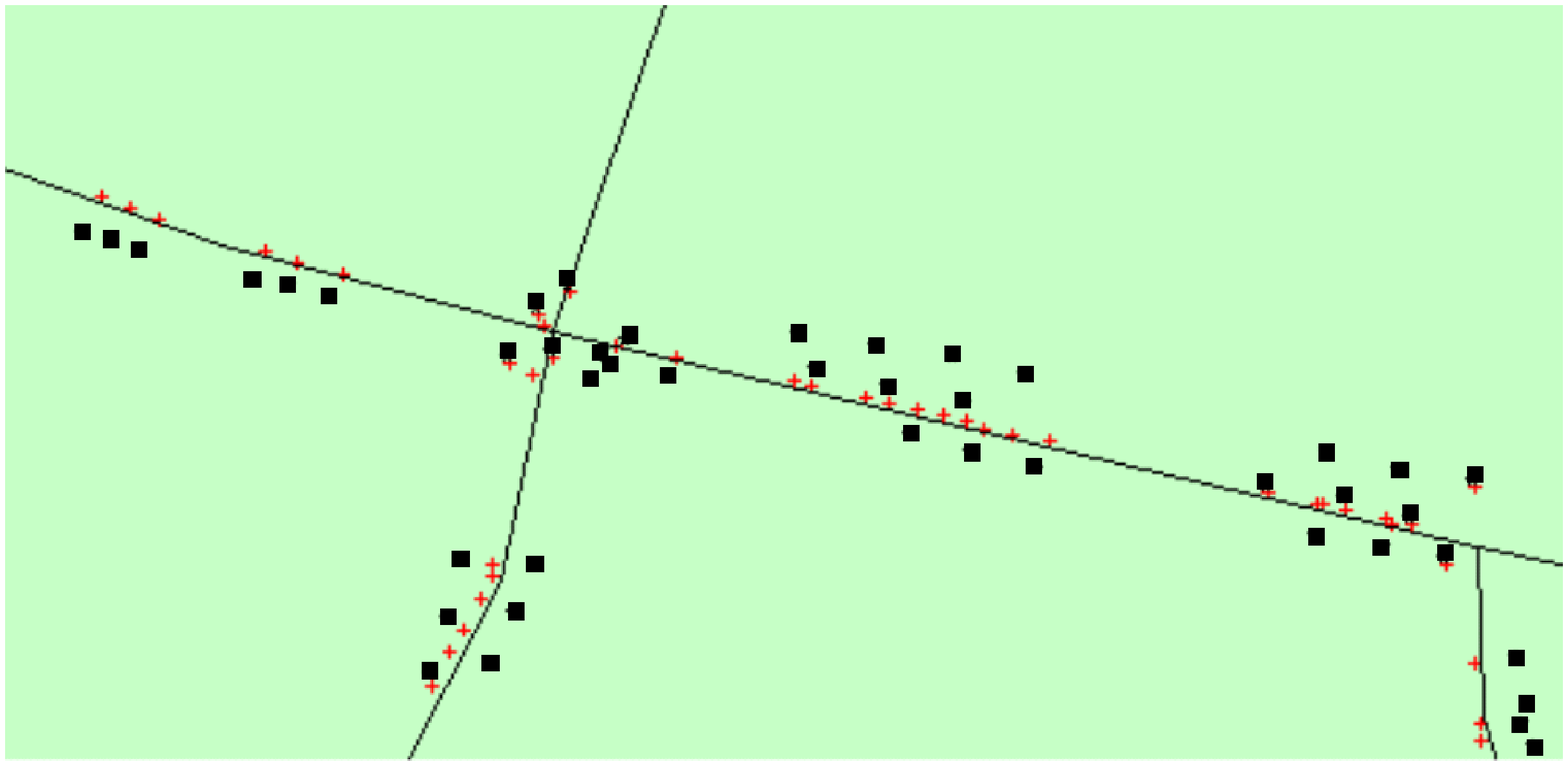}
\end{center}
\caption{Plot of Example Car Data}
\label{fig:mapmach}
\end{figure}

Figure~\ref{fig:mapmach} shows an example of data plot. The figure
uses a {\small ``$\blacksquare$''} to represent a pair of $X$ and $Y$
coordinates obtained from the GPS receiver, and it uses ``+'' symbols
for positions mapped to the roads. One should note that when the car
is near a crossroads, the coordinates are not mapped to the road.

\subsection{Privacy Protection}

With a complete GPS log for a car, it is fairly straightforward to
locate the residence of the driver(s) and thus to identify the
driver(s). To afford the drivers some measure of privacy, we have
applied the procedure described next to the GPS log data.

Specifically, we remove log entries with GPS positions that are close
to the residence of the driver. To do so, the following steps were
applied to each log data file.

\begin{enumerate}
\item The entire area within which the car has been moving is divided
  into squares of size 100m$\times$100m.
\item For each square, we count the number of GPS coordinates that
  first appeared (started) in the square after 4:00~a.m.
\item The square with the largest sum is chosen as the square within
  which the residence of the driver(s) lies.
\item To ensure that the ``right'' square is found, we compare
  visually with real positions on the map.
\item Finally, log entries are removed that intersect with a
  2km$\times$2km square that is chosen at random such that its center
  is less than 1 km from the residence of the driver(s).
\end{enumerate}

\begin{figure}[!htb]
\begin{center}
  \includegraphics[angle=0,width=0.7\textwidth]{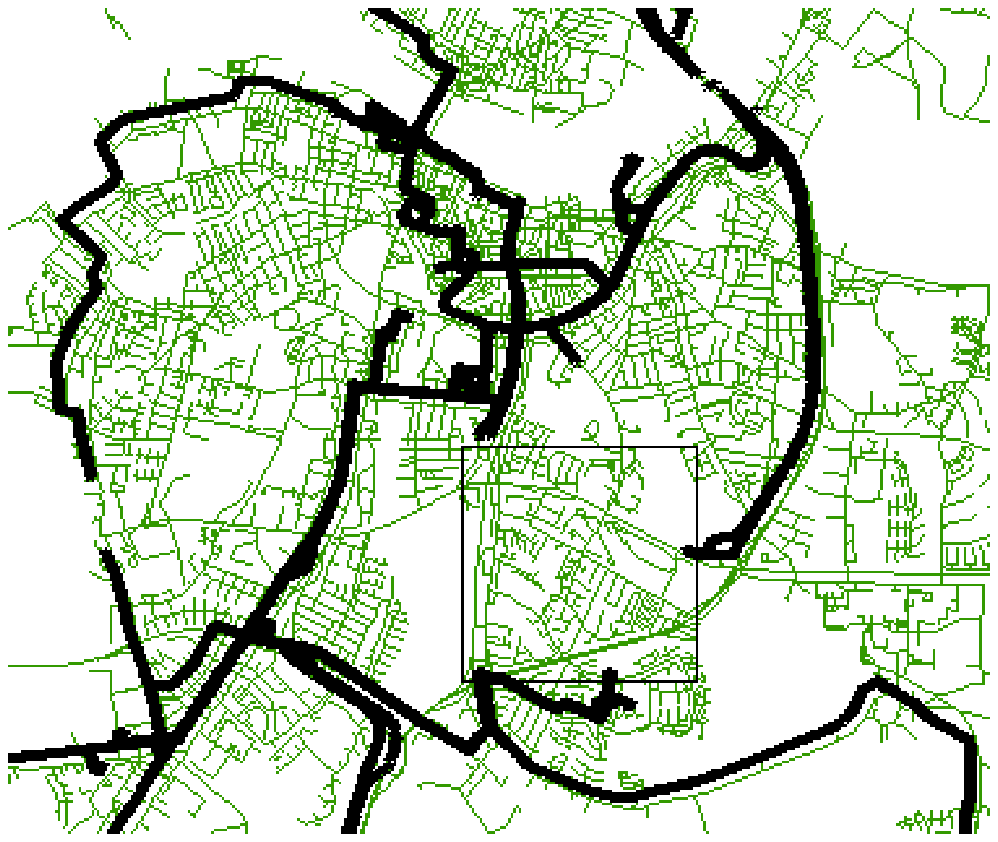}
\end{center}
\caption{Cleaned Car Data}\label{fig:cleaning}
\end{figure}

An example of GPS log data for a car after application of this
procedure is displayed in Figure~\ref{fig:cleaning}.
We use bold lines to represent the GPS coordinates of the car. The
thinly lined rectangle bounds the area close to the residence of the
driver(s). The data inside the rectangle is removed.

\section{Road Network Description}
\label{sec:roadnet}
        
We proceed to describe the road network in which the cars travel. We
first describe the network representation, then describe modifications
we made to the network representation. We have been unable to obtain
permission to distribute this data. The ensuing description serves to
explain better the GPS log data.

\subsection{Road Network Format}

The road network data resides in two files, \emph{road.dat} and
\emph{streetId\_StreetName.txt}. File \emph{road.dat} contains the
road geometry, and its format is given in Table~\ref{d:road.dat}.

\begin{table}[!ht]\center
\begin{tabular}{p{1.7cm}|l}
Attribute & Description \\ \hline
\emph{x\_coord}       & $x$ coordinate of the road segment.\\
\emph{y\_coord}       & $y$ coordinate of the road segment.\\
\emph{street\_code}   & Street code of the road to which the road segment belongs.\\
\emph{kmh}            & Speed allowed on the road segment in kilometers per hour.\\
\emph{unique}         & Not used. \\
\end{tabular}
\caption{Description of File road.dat}
\label{d:road.dat}
\end{table}

A road network is composed of a set of segments. A segment is usually
a part of a road that lies in-between a pair of consecutive
intersections situated along the road. A segment is defined by a
sequence of coordinates. Streets are numbered and are composed of
several road segments.
In file \emph{road.dat}, a segment is thus represented by a set of
entries. The value ``$-9$'' of attribute \emph{street\_code} in an entry
indicates that the entry contains the last coordinate of a segment.
Other values of this attribute identify the street to which the
segment belongs.
A small sample of entries from file \emph{Road.dat} is shown in
Table~\ref{sample:road.dat}.
\begin{wrapfigure}{l}[0pt]{200pt}
\vspace*{-18pt}
\footnotesize\center
\begin{tabular}{@{\hspace{0.2cm}} l @{\hspace{0.2cm}} l @{\hspace{0.2cm}} l @{\hspace{0.2cm}} l @{\hspace{0.2cm}} l }
\emph{x\_coord} & \emph{y\_coord} & \emph{street\_code} & \emph{kmh} & \emph{unique} \\
\hline
  55430572  & 632455870 &7486  &50    &23 \\
  55430979  & 632457914 &7486  &50    &23 \\
  55431749  & 632458306 &7486  &50    &23 \\
  55449649  & 632456885 &-9    &0     &0 \\
  55419427  & 632454790 &6607  &50    &23 \\
  55417961  & 632455407 &6607  &50    &23  \\
  55416386  & 632455047 &-9    &0     &0  \\
  55416386  & 632455047 &6607  &50    &23  \\
  55414107  & 632454593 &6607  &50    &23 \\
  55410829  & 632454465 &-9    &0     &0
\end{tabular}
\tabcaption{Entries from File road.dat}
\label{sample:road.dat}
\end{wrapfigure}
The table contains three segments. The first segment is a polyline
described by four coordinates; it has street code 7486. The next two
segments are composed of polylines described by three coordinates each,
and they belong to the street with code 6607.

\begin{figure}[!!!!!!!h]
\begin{center}
  \includegraphics[angle=0,width=0.9\textwidth]{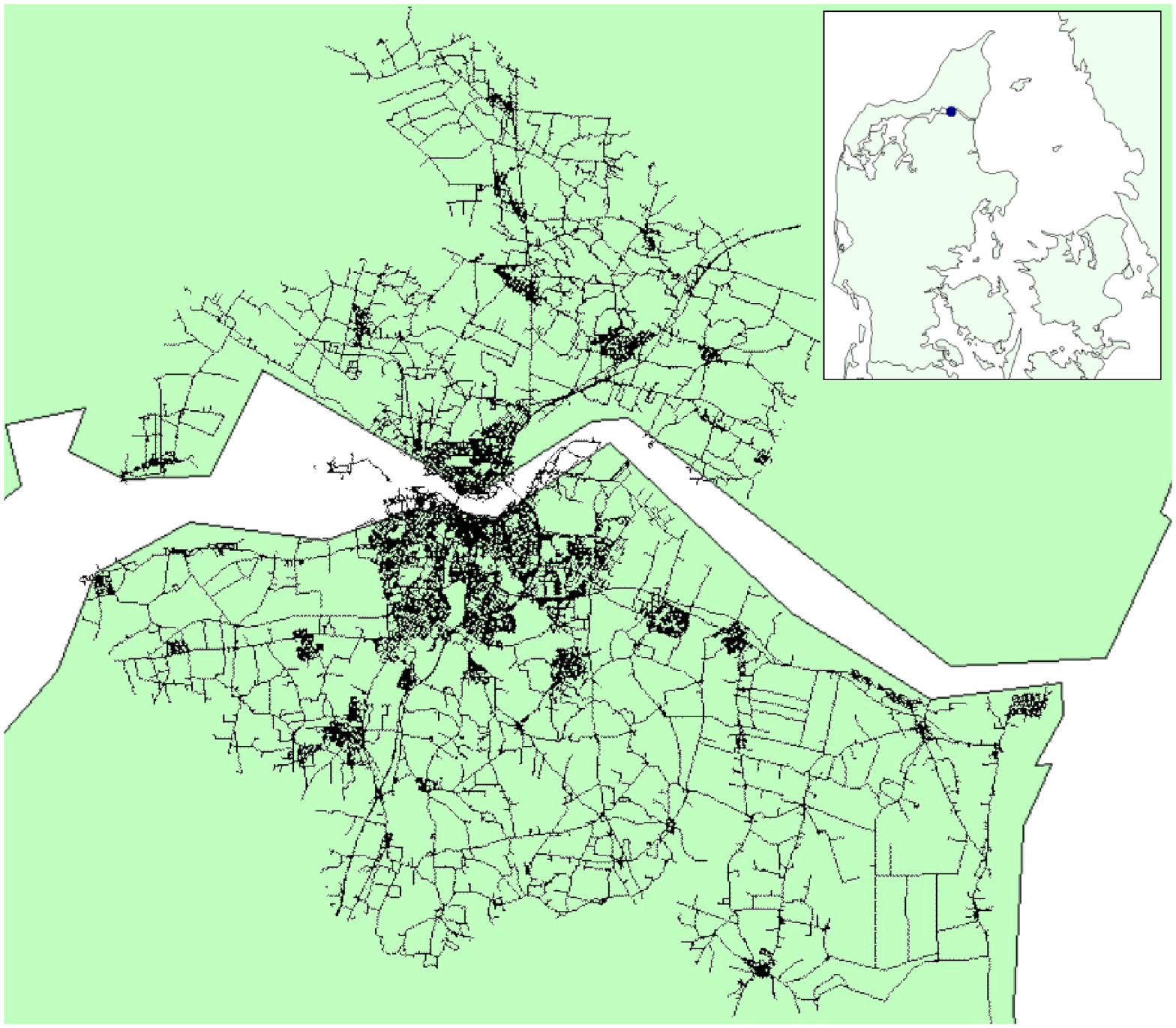}
\end{center}
\caption{Road Map}\label{fig:roadmap}
\end{figure}

The entire road network is shown in Figure~\ref{fig:roadmap}.
The white-and-green background has been included for illustrative
purposes only---it is not part of the road network data. White areas
indicate water, while green areas include land. To the right in the
figure, we thus see part of the East Coast of Northern Jutland. The
white polygon that bisects the figure represents a very rough
approximation of the fjord Limfjorden (e.g., note that an island is
not included).

File \emph{streetId\_StreetName.txt} contains the actual names of the
streets.  Its structure is described in Table~\ref{fig:idname}(a), and
an example of entries from the file is shown in
Table~\ref{fig:idname}(b).

\begin{wrapfigure}{l}[0pt]{325pt}
  \vspace*{-12pt}
  \begin{center}
    \subfigure[Format]{
\begin{tabular}{p{1.7cm}|l}
Column          & Description \\
\hline
\emph{street\_code}  & The code of the street. \\
\emph{street\_name}  & The name of the street.
\end{tabular}
    }
    \hspace{5pt}
    \subfigure[Example Entries]{
\footnotesize
\begin{tabular}{@{\hspace{0.2cm}} l @{\hspace{0.2cm}} l }
 \emph{street\_code} & \emph{street\_name} \\
\hline
-9      & NULL \\
0068    & ABELSVEJ \\
0073    & ABILDGÅRDSVEJ \\
0078    & ABSALONSGADE
\end{tabular}
    }
  \end{center}
  \vspace*{-18pt}
\tabcaption{Street Names in File \emph{streetId\_StreetName.txt}}
\label{fig:idname}
\end{wrapfigure}

\subsection{Road Network Modifications} 
\label{sec:fix}

The road network data was created some time before the GPS log data
were collected. As the road network evolves continually, the road
network data does not quite correspond to the road network in which
the cars actually traveled during the GPS log data collection.

Consequently, there are differences between the roads on which GPS
positions were recorded and the digital road network.  This has led us
to making some modifications of the road network data for some of the
most-used roads.  We have also split some segments that spanned more
than two intersections. This was done in order to ensure that each
road segment is delimited by two consecutive road intersections. The
modified road network is stored in the file \emph{road\_modified.dat}
and has the same format as file \emph{road.dat}. Note that here, the
last two digits of \emph{x\_coord} and \emph{y\_coord } are rounded.

\section{Terms of Usage and Download Information}
\label{sec:download}

The INFATI data can be used free of charge for non-commercial research
purposes. Commercial use is not allowed.  The data can be downloaded
via $<$http:www.cs.auc.dk/TimeCenter/software.htm$>$.  Here, the
following files may be found.

\begin{center}
\begin{tabular}{l|p{12cm}}
File name   & Description \\ \hline
gpsData.zip &  Archive with GPS log data files as described in
Section~\ref{sec:gps}. \\

readme.txt & Short description of the archives and files. \\
TC-TR-79 & This article.
\end{tabular}
\end{center}

\section*{Acknowledgments}

A number of people and organizations contributed to enabling the
publication of this document and the INFATI data.

We thank the members of the INFATI project. The M-Track project,
sponsored by the Electronics and Telecommunications Research
Institute, South Korea, funded in part the production of this
document.  Specifically, in addition to Stardas Pakalnis, it funded
Linas Bukauskas and Alminas \v{C}ivilis, who provided helpful comments
on an earlier draft.
Funding was also received from the European Commission through
contract number IST-2001-32645.

We thank the anonymous drivers, who gave us permission to publish
their GPS data. The road network representation used for map matching
and described in Section~\ref{sec:roadnet} was provided by COWI.

\end{document}